\documentclass[prl,aps,preprintnumbers,nofootinbib,twocolumn,groupedaddress,superscriptaddress]{revtex4-2}

\usepackage[utf8]{inputenc}
\usepackage{graphicx}
\usepackage{slashed}
\usepackage{color}
\usepackage{amsmath}
\usepackage{amssymb}
\usepackage{nameref}
\usepackage{ulem} 

\usepackage{multirow}
\usepackage{tabularx}

\def\lc{\mathrm{LC}}
\def\bbbar{\ensuremath{b\kern1pt\bar{b}}}
\def\wbb{\ensuremath{\operatorname{W}\kern-1pt \bbbar}}
\def\Pp{\ensuremath{\operatorname{p}\kern-1pt}}
\def\PW{\ensuremath{\operatorname{W}\kern-1pt}}
\def\PZ{\ensuremath{\operatorname{Z}\kern-1pt}}
\def\PH{\ensuremath{\operatorname{H}\kern-1pt}}
\def\RT{\ensuremath{\text{T}}}
\def\cF{\mathcal{F}}
\def\cV{\mathcal{V}}
\def\cM{\mathcal{M}}
\def\cD{\mathcal{D}}
\def\cO{\mathcal{O}}
\def\gev{\mathrm{GeV}}

\def\Kinc{\mathcal{K}_\mathrm{inc}}
\def\Kexc{\mathcal{K}_\mathrm{exc}}
\def\lo{\mathrm{LO}}
\def\nlo{\mathrm{NLO}}
\def\nnlo{\mathrm{NNLO}}

\definecolor{mypink}{RGB}{219, 48, 122}
\definecolor{mygreen}{rgb}{0,0.7,0}

\begin{document}

\newcolumntype{C}[1]{>{\hsize=#1\hsize\centering\arraybackslash}X}

\preprint{CAVENDISH-HEP-22/05}

\title{NNLO QCD corrections to $\wbb$ production at the LHC}

\author{Heribertus Bayu Hartanto}
\email{hbhartanto@hep.phy.cam.ac.uk}
\affiliation{
Cavendish Laboratory, University of Cambridge, Cambridge CB3 0HE, United Kingdom
}

\author{Rene Poncelet}
\email{poncelet@hep.phy.cam.ac.uk}
\affiliation{
Cavendish Laboratory, University of Cambridge, Cambridge CB3 0HE, United Kingdom
}

\author{Andrei Popescu}
\email{popescu@hep.phy.cam.ac.uk}
\affiliation{
Cavendish Laboratory, University of Cambridge, Cambridge CB3 0HE, United Kingdom
}

\author{Simone Zoia}
\email{simone.zoia@unito.it}
\affiliation{
Dipartimento di Fisica and Arnold-Regge Center, Università di Torino, and INFN, Sezione di
Torino, Via P. Giuria 1, I-10125 Torino, Italy
}

\date{\today}

\begin{abstract}
We compute theoretical predictions for the production of a W-boson in association with a bottom-quark pair at hadron colliders at
next-to-next-to-leading order (NNLO) in QCD, including the leptonic decay of the W-boson, while treating
the bottom quark as massless. This calculation constitutes the very first $2\to3$ process with a massive external particle to be studied at such a perturbative order.
We derive an analytic expression for the required two-loop five-particle amplitudes in the leading colour approximation employing finite-field methods.
Numerical results for the cross section and differential distributions are presented for the Large Hadron Collider at $\sqrt{s}=8$ TeV.
We observe an improvement of the perturbative convergence for the inclusive case and for the prediction with a jet veto upon the inclusion of the NNLO QCD corrections.
\end{abstract}

\maketitle

\section{Introduction \label{sec:intro}}

Studying vector boson production in association with multi-jet final states at the Large Hadron Collider (LHC) offers a wide variety of interesting phenomenological explorations.
In particular, the production of a W-boson in association with bottom quark ($b$) jets is very interesting both from experimental and theoretical perspectives.
It is crucial to scrutinise experimental signatures for both $\PW+1b$ jet and $\PW+2b$ jets, in order to test our knowledge of the strong interaction at high energies
and improve our modelling of bottom-quark jets at the LHC.
The cross sections for both the signatures have been measured at the Tevatron~\cite{D0:2004prj,D0:2012qt} 
and LHC~\cite{ATLAS:2011jbx,ATLAS:2013gjg,CMS:2013xis,CMS:2016eha}.
While the $\PW+1b$ jet signature provides a fundamental probe of the bottom-quark parton distribution functions (PDFs), the $\PW+2b$ jets final state constitutes an irreducible background to many
important reactions studied at the LHC, such as the Higgs-strahlung process ($\Pp\Pp\to \PW\PH\ (\PH\to \bbbar)$) and single top production ($\Pp\Pp\to \bar{b}\,t\ (t\to \PW b)$),
as well as many beyond the Standard Model (BSM) searches.
Moreover, the $\PW+b$ jets processes are interesting from a theoretical point of view as they are a perfect testing ground to study the different ways of treating the $b$-quark. In particular, the choice of whether to take its mass and presence in the PDF into account, leads to two disparate computational schemes: the four- (4FS) and five-flavour number schemes (5FS).

In this letter we compute the NNLO QCD corrections to W-boson production in association with a bottom-quark pair, which we henceforth call $\wbb$,
including the leptonic decay of the W-boson ($\PW\to\ell\nu)$. We work in 5FS, thus treating the bottom quark, and additionally the charged leptons, as massless particles. 
They contribute to the $\PW+2b$ jets signature, as well as $\PW+1b$ jet production when at least one $b$-jet is tagged.
Extensive studies of $\wbb$ production at NLO QCD accuracy~\cite{Ellis:1998fv,FebresCordero:2006nvf,FebresCordero:2009xzo,Badger:2010mg,
Oleari:2011ey,Frederix:2011qg} indicate a poor perturbative behaviour at such order (i.e.\ the corrections are large and
the scale uncertainties do not improve with respect to the leading order predictions for inclusive final state) due to the opening of the $qg$-initiated channel.
Several efforts to assess corrections beyond NLO were done by including additional jet radiations~\cite{Luisoni:2015mpa,Anger:2017glm}.
It is clear that a fully-fledged NNLO QCD prediction is mandatory to improve the perturbative convergence of $\wbb$ production.

Improvement of theoretical precision is also a critical component of the progress in Particle Physics, as we enter the precision LHC era with the upcoming Run 3 and high-luminosity phases.
We have seen spectacular breakthroughs in perturbative QCD calculations in the recent years, with a number of $2\to 3$ processes computed at NNLO QCD accuracy for fully massless
final states~\cite{Chawdhry:2019bji, Kallweit:2020gcp, Chawdhry:2021hkp, Czakon:2021mjy, Badger:2021ohm, Chen:2022ktf}.
This success stems from both the advancements in the scattering amplitude computations and the developments of NNLO subtraction schemes.

The analytic computation of the required two-loop five-particle amplitudes is one of the main bottlenecks towards achieving NNLO QCD accuracy for $2\to 3$ processes.
However, the progress for five-particle processes with a single massive external particle has been spectacular recently.
All planar two-loop five-particle Feynman integrals are now available analytically~\cite{Papadopoulos:2015jft,Abreu:2020jxa,Canko:2020ylt,Syrrakos:2020kba}
in terms of bases of special functions which substantially simplify computation of the finite remainders,
and allow for an extremely efficient numerical evaluation~\cite{Badger:2021nhg,Chicherin:2021dyp}.
Partial results for the non-planar integral families have also become available recently~\cite{Papadopoulos:2019iam,Abreu:2021smk,Kardos:2022tpo}. This progress resulted in
a number of two-loop amplitudes computed at leading colour~\cite{Badger:2021nhg,Badger:2021ega,Abreu:2021asb,Badger:2022ncb}.
In this work we derive the analytic form of the leading colour two-loop amplitude contributing to $\PW\,(\to \ell\nu)\,\bbbar$ production, and employ it to compute a number of observables for this process at NNLO in QCD.
Our computation marks a significant precision-calculation milestone, since it represents the very first prediction to be derived for a $2\to 3$ process involving a massive final state.


This letter is structured as follows. We begin by discussing the derivation of the required two-loop matrix elements and the tools used to perform the cross section computation, and then present phenomenological results for the cross section and a number of interesting differential distributions.

\section{Calculation  \label{sec:calculation}}
We consider $\Pp\Pp \to \ell^+\nu\,\bbbar$ production (with $\ell=e$ or $\mu$) up to $\cO(\alpha^2 \alpha_s^4)$.
The calculation has been performed within the \textsc{Stripper} framework, a \textsc{C++} implementation of the four-dimensional formulation of the sector-improved residue
subtraction scheme~\cite{Czakon:2010td,Czakon:2014oma,Czakon:2019tmo}.
The tree-level matrix elements are supplied by the \textsc{AvH} library~\cite{Bury:2015dla}, while the one-loop matrix elements are provided by the \textsc{OpenLoops} package~\cite{Buccioni:2017yxi,Buccioni:2019sur}.
We compute the double virtual contribution $\cV^{(2)}$ in the leading colour approximation for
\begin{equation}
u(p_1) + \bar{d}(p_2) \to b(p_3) + \bar{b}(p_4) + \ell^+(p_5) + \nu(p_6) \,.
\end{equation}
It consists of the (colour and helicity summed) two-loop and one-loop squared matrix elements,
\begin{equation}
\cV^{(2)} = \sum_{\text{col.}}  \sum_{\text{hel.}} \left\{ 2 \, \mathrm{Re}\big[ M^{(0)*} \cF^{(2)} \big] + \big|\cF^{(1)}\big|^2 \right\} \,,
\end{equation}
where $M^{(0)}$ is the tree-level amplitude, and $\cF^{(L)}$ is the $L$-loop finite remainder.
We decompose $\cV^{(2)}$ at leading colour into
\begin{align}
\cV^{(2)}_\lc & = \cV^{(2),1} + \frac{n_f}{N_c} \cV^{(2),n_f} + \frac{n_f^2}{N_c^2} \cV^{(2),n_f^2} \,,
\end{align}
where $n_f$ is the number of massless closed fermion loops.
We note that the leading-colour approximation is only enforced in the scale-independent part of the double virtual contribution,
\begin{align}\label{eq:vv}
\cV^{(2)}(\mu_R^2) & = \cV^{(2)}_{\lc}(s_{12}) + \sum_{i=1}^{4} c_i \ln^i \bigg( \frac{\mu_R^2 }{s_{12}} \bigg)\,,
\end{align}
where $s_{ij} = (p_i + p_j)^2$, and the kinematic-dependent coefficients $c_i$ are expressed in terms of full colour lower-order matrix elements.

The analytic computation of the two-loop amplitude follows closely Ref.~\cite{Badger:2021nhg}, with modifications implemented to incorporate the decay of the W-boson.
Since the QCD corrections do not apply to the $\PW \to \ell\nu$ decay, we can
separate the 6-point squared amplitude $\cM^{(2)}_6$ into the product of the 5-point W-production squared amplitude $\cM^{(2)}_{5\mu\nu}$
and the leptonic tensor $\cD^{\mu\nu}$,
\begin{align}
\cM_{6}^{(2)} = \cM_{5\mu\nu}^{(2)} \; \cD^{\mu\nu} \; |P(s_{56})|^2 \,,
\end{align}
where $P(s) = 1/(s-M_{\PW}^2 + i M_{\PW} \Gamma_{\PW})$ is the W-boson propagator factor.
We perform tensor decomposition on the 5-point W-production squared amplitude,
\begin{align}
\cM_{5}^{(2)\mu\nu} = \sum_{i=1}^{16} a^{(2)}_{i} v_{i}^{\mu\nu}\,,
\label{eq:5ptdecomposition}
\end{align}
using $\lbrace p_1,p_2,p_3,p_5+p_6\rbrace$ as the spanning basis to build the $v_i^{\mu\nu}$ basis tensors \cite{Chen:2019wyb,Peraro:2020sfm}.
The coefficients $a^{(2)}_{i}$ can be determined by contracting Eq.~\eqref{eq:5ptdecomposition} with $v_{i\mu\nu}$ and inverting the resulting linear system of equations.
The analytic form of the \textit{contracted} squared amplitudes $v_{i\mu\nu}\cM_{5}^{(2)\mu\nu}$ was derived using finite-field reconstruction methods within the \textsc{FiniteFlow}
framework~\cite{Peraro:2016wsq,Peraro:2019svx}. 
We expressed them in terms of the special functions of Ref.~\cite{Chicherin:2021dyp} and rational coefficients, 
which we simplified using \textsc{MultivariateApart}~\cite{Heller:2021qkz} and  \textsc{Singular}~\cite{DGPS}. 
We further implemented these amplitudes in \textsc{C++} for a fast numerical evaluation. We evaluate the special functions using the \textsc{PentagonFunction++} library~\cite{Chicherin:2021dyp}.
Our analytic result is validated numerically against the $\PW+4$ quarks helicity amplitudes derived in Ref.~\cite{Abreu:2021asb} at the level of the helicity-summed squared finite remainder.
The complete analytic expression is included in ancillary files.

\section{Phenomenology \label{sec:results}}

We present numerical results for the LHC center-of-mass energy $\sqrt{s} = 8$ TeV, focusing on the $\PW^+(\to \ell^+ \nu)\,\bbbar$ final state.
The Standard Model input parameters are
\begin{align*}
M_{\PW} &= 80.351972 \; \gev, & \Gamma_{\PW} = 2.0842989 \; \gev, \nonumber \\
M_{\PZ} &= 91.153481 \; \gev, & \Gamma_{\PZ} = 2.4942665 \; \gev, \\
G_F &= 1.16638 \cdot 10^{-5} \;\gev^{-2}, & \nonumber
\end{align*}
from which the electromagnetic coupling $\alpha$ can be derived within the $G_\mu$ scheme.
We assume a diagonal CKM matrix and employ the \verb=NNPDF31_as_0118= PDF sets~\cite{NNPDF:2017mvq} with its perturbative order matching that of the corresponding calculations.
Since we treat the bottom quark as massless, we need a flavour-sensitive jet algorithm
to define the flavoured jets in an infrared-safe way. The partons are clustered into a jet using the flavour-$k_\RT$ jet algorithm~\cite{Banfi:2006hf} with $R=0.5$. 
The jets (including $b$-jets) and charged leptons are required to fulfill the following event-selection criteria~\cite{CMS:2016eha}:
\begin{equation}
p_{\RT,\ell} > 30 \;\gev,\;\; |\eta_{\ell}| < 2.1, \;\; p_{\RT,j} > 25 \;\gev, \;\; |\eta_{j}| < 2.4. \nonumber
\end{equation}
The renormalisation and factorisation scales are set to a common value $\mu_R = \mu_F = H_\RT$,
with
\begin{equation}
H_\RT = E_\RT(\ell\nu) + p_\RT(b_1) + p_\RT(b_2)\,,
\end{equation}
where $b_1$, $b_2$ are correspondingly the hardest and second hardest $b$-flavoured (either $b$ or $\bar b$) jets.
Unless otherwise specified, the scale uncertainties are obtained using the 7-point scale variation, where $\mu_R$ and $\mu_F$ are varied by a factor of 2 around $H_\RT$, while satisfying the $1/2 \le \mu_R/\mu_F \le 2$ constraint.

Based on the number of jets  required in the final states, we can define the following configurations for the NLO and NNLO predictions:
\begin{itemize}
\item inclusive (inc): at least 2 $b$-jets;
\item exclusive (exc): exactly 2 $b$-jets and no other jets.
\end{itemize}
Na\"ive scale variation of the exclusive prediction may lead to an underestimation of the scale uncertainties~\cite{Stewart:2011cf}.
Hence, for the exclusive configuration, we use also the uncorrelated prescription of Ref.~\cite{Stewart:2011cf}, in addition to the 7-point scale variation.

\renewcommand{\arraystretch}{1.5}
\begin{table}[t!]
\centering
\begin{tabularx}{0.475\textwidth}{|C{0.6}|C{1.4}|C{0.6}|C{1.8}|C{0.6}|}
\hline
  & inclusive [fb] & $\Kinc$ & exclusive [fb] & $\Kexc$ \\
\hline
$\sigma_\lo$
               & $213.2(1)^{+21.4\%}_{-16.1\%}$
               & -
               & $213.2(1)^{+21.4\%}_{-16.1\%}$
               & -
               \\
$\sigma_\nlo$
               & $362.0(6)^{+13.7\%}_{-11.4\%}$
               & 1.7
               & $249.8(4)^{+3.9 (+27)\%}_{-6.0 (-19)\%}$
               & 1.17
               \\
$\sigma_\nnlo$
               & $445(5)^{+6.7\%}_{-7.0\%}$
               & 1.23
               & $267(3)^{+1.8 (+11)\%}_{-2.5 (-11)\%}$
               & 1.067
               \\
\hline
\end{tabularx}
\caption{\label{tab:xsection}
Fiducial cross sections for $\Pp\Pp\to \ell^+\nu b \bar{b}$ production at the LHC with $\sqrt{s}=8$ TeV at LO, NLO and NNLO for both inclusive (inc) and exclusive (exc) final states.
The corresponding $\mathcal{K}$ factor is defined as $\mathcal{K} = \sigma_{\mathrm{N^{(n)}LO}}/\sigma_{\mathrm{N^{(n-1)}LO}}$.
The statistical errors are shown for the central predictions.
Scale uncertainties for the exclusive predictions are provided using both the standard 7-point scale variation and uncorrelated prescription of Ref.~\cite{Stewart:2011cf}.
The latter is quoted inside  parentheses in the error estimates. 
}
\end{table}

In Table~\ref{tab:xsection}, we present numerical results for the fiducial cross section for the inclusive and exclusive configurations at different perturbative orders.
As observed in the previous studies~\cite{FebresCordero:2009xzo,Anger:2017glm}, 
the NLO QCD corrections are large in the case of the inclusive phase space. In our calculation this amounts to about $70\%$ corrections.
The jet veto in the exclusive selection reduces the NLO QCD corrections to a moderate $17\%$.
A similar observation holds at NNLO QCD, where we find a positive correction of $23\%$ in the inclusive and $6.7\%$ in the exclusive case.
The NNLO QCD corrections are smaller than the NLO QCD corrections in both cases indicating perturbative convergence.
In that respect, by using the scale dependence as the canonical way to estimate the uncertainties from missing higher orders, we conclude that theoretical uncertainty reduces with inclusion of higher order terms.
However, for the inclusive phase space, the NLO corrections are significantly larger than the LO scale dependence.
The situation at NNLO QCD slightly improves, but the corrections are still only barely covered by the NLO scale band.
For the exclusive case, the NLO corrections are within the LO band, however the estimated uncertainty from the 7-point scale variation is comparatively small, only $5\%$.
 The NNLO corrections here are also smaller, but are well outside the NLO scale uncertainty, indicating that the NLO scale dependence is underestimated.
This motivates the alternative prescription of Ref.~\cite{Stewart:2011cf} to estimate theory uncertainties, taking into account the jet veto effect.
The uncertainties resulting from this prescription are shown in the parentheses and are significantly larger.
The higher order corrections fall well within the uncertainty bands, implying that this method is more reliable, but also quite conservative.

The double virtual corrections, which have been included only in the leading colour approximation, deserve an additional comment.
For the inclusive setup, we find that the contribution of Eq.~\eqref{eq:vv} to the cross section is about $5\%$.
In the exclusive case, the Born configurations are unaffected by the jet veto, but a fraction of the hard radiative corrections are suppressed.
This leads to an enhancement of the sensitivity to the double virtual matrix element, which contributes $\sim\!10\%$ of the fiducial cross section in this case.
The na\"ive expectation for the subleading colour effects is that they are about $10\%$ of the double virtual matrix element, 
implying that potential corrections to the fiducial cross section would be about $1\%$ ($0.5\%$) for the exclusive (inclusive) case.
\begin{figure}[!hb]
  \centering
  \includegraphics[width=0.99\columnwidth]{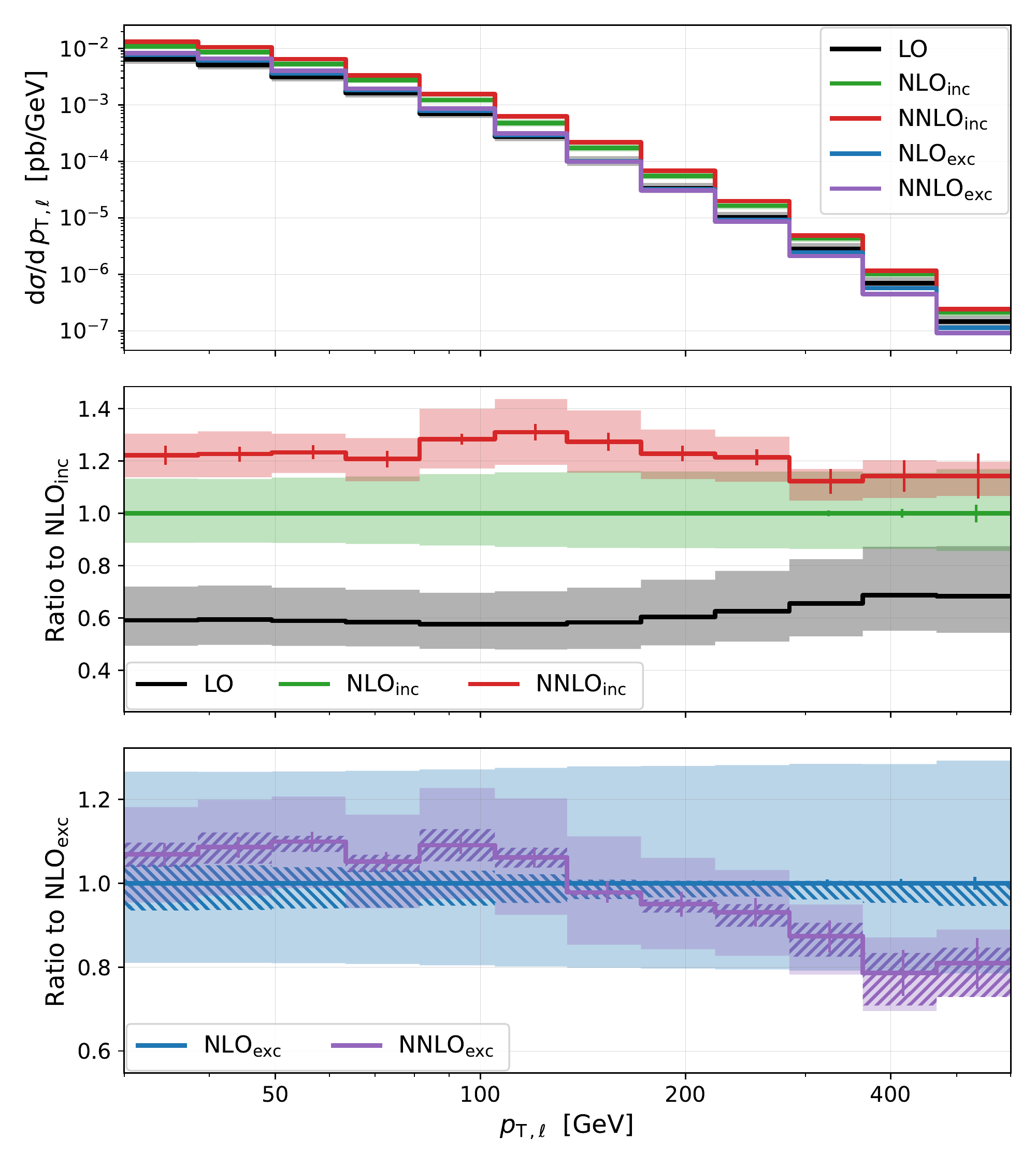}
  \caption{The charged lepton's transverse momentum distribution. The upper panel shows the absolute predictions for the inclusive and exclusive selection at different perturbative orders. The middle panel shows the inclusive cross sections as a ratio with respect to the central NLO prediction, with the coloured bands indicating the 7-point scale variation. The lower panel shows the same ratio for the exclusive configuration. Here, the coloured bands correspond to the decorrelated scale variation, and the hashed bands to the standard 7-point variation. The vertical bars indicate the statistical uncertainty.}
  \label{fig:10_pTl}
\end{figure}

Turning to the differential distributions, we present the transverse momentum of the charged lepton, $p_{\RT,\ell}$, in Fig.~\ref{fig:10_pTl}, for the inclusive, as well as exclusive, phase space selection.
Focusing on the perturbative convergence of the spectrum, we can draw similar conclusions as for the fiducial cross section.
In the inclusive case, we find sizeable NNLO QCD corrections of $\sim\!20\%$, which are barely contained in the NLO uncertainty.
The corrections have a tendency to increase at higher energies, being the largest around $p_{\RT,\ell}\approx 100$ GeV, similarly to the NLO corrections.
For the exclusive phase space, we find positive corrections of about $7\%$ for low $p_{\RT,\ell}$, and negative corrections of order $\sim\!10\%$ for $p_{\RT,\ell}>100$ GeV.
Again, we observe that the decorrelated prescription to estimate the uncertainty is more reliable.

\begin{figure}[!t]
  \centering
  \includegraphics[width=0.99\columnwidth]{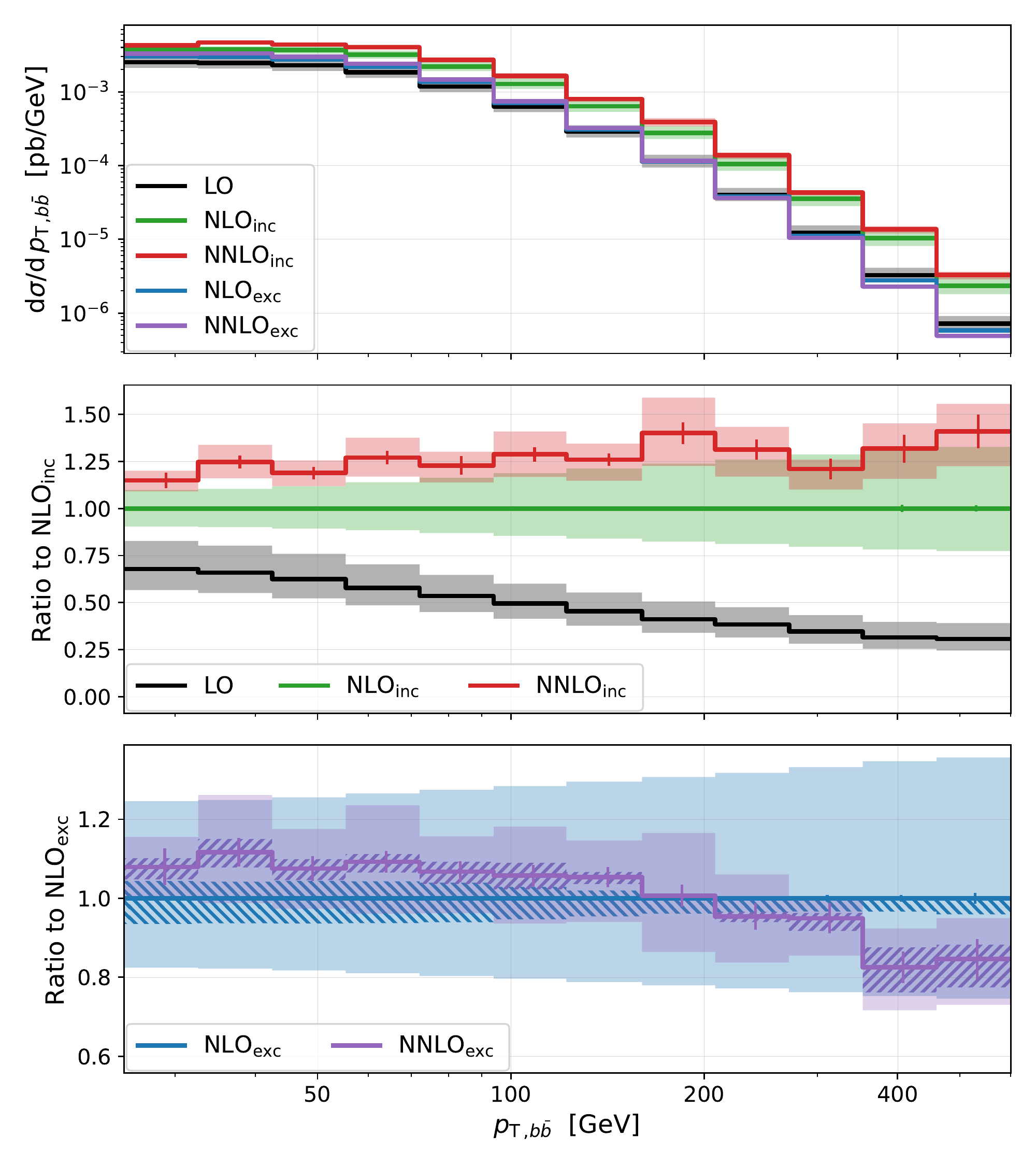}
  \caption{Transverse momentum distribution of the $b\bar{b}$ system. Same layout as in Fig.~\ref{fig:10_pTl}.}
  \label{fig:6_pTbb}
\end{figure}

The next two distributions characterise the $\bbbar$ system.
In Fig.~\ref{fig:6_pTbb}, we show the transverse momentum of the $\bbbar$ system, $p_{\RT,b\bar{b}}$.
In terms of perturbative corrections we find a similar trend as for the charged lepton transverse momentum.
Additionally, the absolute distributions highlight that the inclusive spectrum is, in general, harder than the exclusive case, confirming the intuition that the jet veto suppresses additional large transverse momentum jets.
In the case of exclusive phase space, this differential distribution can be understood as a proxy for the W transverse momentum.

\begin{figure}[!t]
  \centering
  \includegraphics[width=0.99\columnwidth]{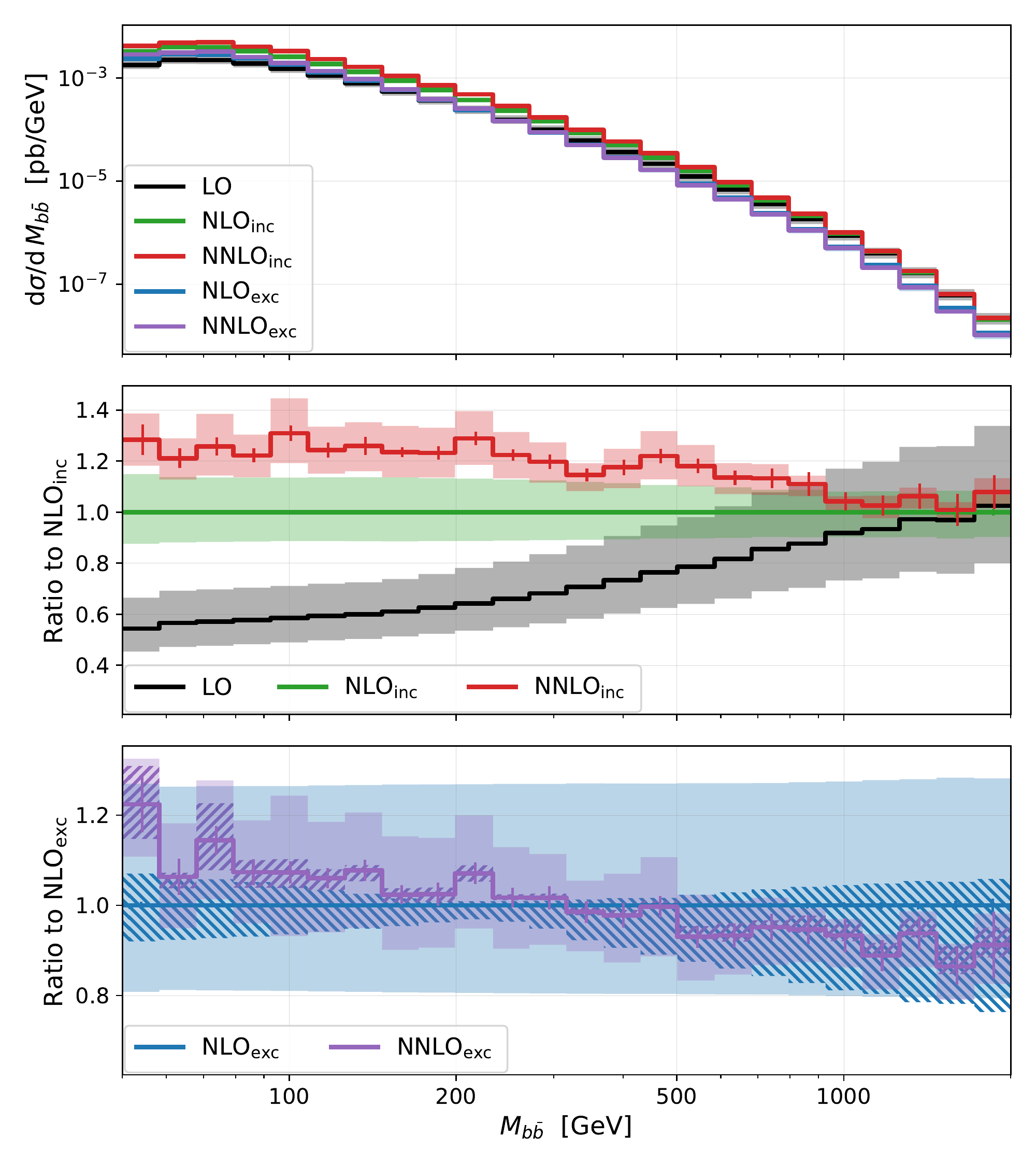}
  \caption{Invariant mass distribution of the $b\bar{b}$ system. Same layout as in Fig.~\ref{fig:10_pTl}.}
  \label{fig:4_Mbb}
\end{figure}

The distribution of the invariant mass of the $\bbbar$ system, $M_{b\bar{b}}$, is shown in Fig.~\ref{fig:4_Mbb}.
This observable is interesting when considering the QCD process $\wbb$ as background to the Higgs-strahlung process $\PW \PH\,(\to b\bar{b})$.
Around the Higgs mass we can see that the NNLO QCD corrections are about $20\%$ in the inclusive selection and only $\sim\!5\%$ in the exclusive case.
By comparing the two prescriptions for estimating the uncertainty, we see that around the Higgs mass the 7-point prescription implies a 2-3 times smaller uncertainty than the decorrelated method.

The reader is invited to find our results for other observables in the auxilliary files to this publication.

\section{Conclusions \label{sec:conclusions}}

We presented fiducial and differential cross sections for the $\wbb$ process at the LHC with 8 TeV center-of-mass energy.
This includes the computation of the double virtual amplitudes in the leading colour approximation with incorporated decay of the W-boson.

We addressed the observation of large NLO QCD corrections in this process, and found that the NNLO QCD corrections are significantly smaller.
We observe a significant reduction of the scale dependence, which indicates perturbative convergence.
We discussed the behaviour of the jet-vetoed cross section, which exhibits much smaller corrections but suffers from accidental cancellations in the scale dependence, rendering the theory uncertainty estimates from canonical scale variation unreliable.
At NNLO accuracy we validated the alternative prescription of Ref.~\cite{Stewart:2011cf} for estimating the theory uncertainties.

This work constitutes the first NNLO QCD calculation for a $2\to3$ process including a massive final state particle.
Studying this class of processes at such accuracy is of the utmost importance for the physics programme of the LHC.
However, the steep requirements for amplitudes involving many loops and high multiplicities used to put them beyond the reach of computational capabilities. 
Our results demonstrate that the door of precision phenomenology is finally open for these processes as well.


\begin{acknowledgments}

The authors would like to thank Micha\l{} Czakon for making the \textsc{Stripper}
library available to us, Vasily Sotnikov for help in the comparison with the results of Ref.~\cite{Abreu:2021asb}, and Simon Badger and Alexander Mitov for many inspiring discussions and useful comments on the draft.
This project received funding from the European Union's Horizon 2020 research and innovation programmes
\textit{New level of theoretical precision for LHC Run 2 and beyond} (grant agreement No 683211),
and \textit{High precision multi-jet dynamics at the LHC} (grant agreement No 772009).
HBH was partially supported by STFC consolidated HEP theory grant ST/T000694/1.
SZ gratefully acknowledges the computing resources provided by the Max Planck Institute for Physics and by the Max Planck Computing \& Data Facility.
AP is also supported by the Cambridge Trust and Trinity College Cambridge.
RP acknowledges the support from the Leverhulme Trust and the Isaac Newton Trust,
as well as the use of the DiRAC Cumulus HPC facility under Grant No.\ PPSP226.
\end{acknowledgments}

\bibliography{wbb_prl}

\begin{thebibliography}{47}%
\makeatletter
\providecommand \@ifxundefined [1]{%
 \@ifx{#1\undefined}
}%
\providecommand \@ifnum [1]{%
 \ifnum #1\expandafter \@firstoftwo
 \else \expandafter \@secondoftwo
 \fi
}%
\providecommand \@ifx [1]{%
 \ifx #1\expandafter \@firstoftwo
 \else \expandafter \@secondoftwo
 \fi
}%
\providecommand \natexlab [1]{#1}%
\providecommand \enquote  [1]{``#1''}%
\providecommand \bibnamefont  [1]{#1}%
\providecommand \bibfnamefont [1]{#1}%
\providecommand \citenamefont [1]{#1}%
\providecommand \href@noop [0]{\@secondoftwo}%
\providecommand \href [0]{\begingroup \@sanitize@url \@href}%
\providecommand \@href[1]{\@@startlink{#1}\@@href}%
\providecommand \@@href[1]{\endgroup#1\@@endlink}%
\providecommand \@sanitize@url [0]{\catcode `\\12\catcode `\$12\catcode
  `\&12\catcode `\#12\catcode `\^12\catcode `\_12\catcode `\%12\relax}%
\providecommand \@@startlink[1]{}%
\providecommand \@@endlink[0]{}%
\providecommand \url  [0]{\begingroup\@sanitize@url \@url }%
\providecommand \@url [1]{\endgroup\@href {#1}{\urlprefix }}%
\providecommand \urlprefix  [0]{URL }%
\providecommand \Eprint [0]{\href }%
\providecommand \doibase [0]{https://doi.org/}%
\providecommand \selectlanguage [0]{\@gobble}%
\providecommand \bibinfo  [0]{\@secondoftwo}%
\providecommand \bibfield  [0]{\@secondoftwo}%
\providecommand \translation [1]{[#1]}%
\providecommand \BibitemOpen [0]{}%
\providecommand \bibitemStop [0]{}%
\providecommand \bibitemNoStop [0]{.\EOS\space}%
\providecommand \EOS [0]{\spacefactor3000\relax}%
\providecommand \BibitemShut  [1]{\csname bibitem#1\endcsname}%
\let\auto@bib@innerbib\@empty
\bibitem [{\citenamefont {Abazov}\ \emph {et~al.}(2005)\citenamefont {Abazov}
  \emph {et~al.}}]{D0:2004prj}%
  \BibitemOpen
  \bibfield  {author} {\bibinfo {author} {\bibfnamefont {V.~M.}\ \bibnamefont
  {Abazov}} \emph {et~al.} (\bibinfo {collaboration} {D0}),\ }\bibfield
  {title} {\bibinfo {title} {{A Search for $W b\bar{b}$ and $W H$ Production in
  $p\bar{p}$ Collisions at $\sqrt{s} = 1.96$ TeV}},\ }\href
  {https://doi.org/10.1103/PhysRevLett.94.091802} {\bibfield  {journal}
  {\bibinfo  {journal} {Phys. Rev. Lett.}\ }\textbf {\bibinfo {volume} {94}},\
  \bibinfo {pages} {091802} (\bibinfo {year} {2005})},\ \Eprint
  {https://arxiv.org/abs/hep-ex/0410062} {arXiv:hep-ex/0410062} \BibitemShut
  {NoStop}%
\bibitem [{\citenamefont {Abazov}\ \emph {et~al.}(2013)\citenamefont {Abazov}
  \emph {et~al.}}]{D0:2012qt}%
  \BibitemOpen
  \bibfield  {author} {\bibinfo {author} {\bibfnamefont {V.~M.}\ \bibnamefont
  {Abazov}} \emph {et~al.} (\bibinfo {collaboration} {D0}),\ }\bibfield
  {title} {\bibinfo {title} {{Measurement of the $p\bar{p} \to W+b+X$
  production cross section at $\sqrt{s}=1.96$ TeV}},\ }\href
  {https://doi.org/10.1016/j.physletb.2012.12.044} {\bibfield  {journal}
  {\bibinfo  {journal} {Phys. Lett. B}\ }\textbf {\bibinfo {volume} {718}},\
  \bibinfo {pages} {1314} (\bibinfo {year} {2013})},\ \Eprint
  {https://arxiv.org/abs/1210.0627} {arXiv:1210.0627 [hep-ex]} \BibitemShut
  {NoStop}%
\bibitem [{\citenamefont {Aad}\ \emph {et~al.}(2012)\citenamefont {Aad} \emph
  {et~al.}}]{ATLAS:2011jbx}%
  \BibitemOpen
  \bibfield  {author} {\bibinfo {author} {\bibfnamefont {G.}~\bibnamefont
  {Aad}} \emph {et~al.} (\bibinfo {collaboration} {ATLAS}),\ }\bibfield
  {title} {\bibinfo {title} {{Measurement of the cross section for the
  production of a $W$ boson in association with $b^-$ jets in $pp$ collisions
  at $\sqrt{s}=7$ TeV with the ATLAS detector}},\ }\href
  {https://doi.org/10.1016/j.physletb.2011.12.046} {\bibfield  {journal}
  {\bibinfo  {journal} {Phys. Lett. B}\ }\textbf {\bibinfo {volume} {707}},\
  \bibinfo {pages} {418} (\bibinfo {year} {2012})},\ \Eprint
  {https://arxiv.org/abs/1109.1470} {arXiv:1109.1470 [hep-ex]} \BibitemShut
  {NoStop}%
\bibitem [{\citenamefont {Aad}\ \emph {et~al.}(2013)\citenamefont {Aad} \emph
  {et~al.}}]{ATLAS:2013gjg}%
  \BibitemOpen
  \bibfield  {author} {\bibinfo {author} {\bibfnamefont {G.}~\bibnamefont
  {Aad}} \emph {et~al.} (\bibinfo {collaboration} {ATLAS}),\ }\bibfield
  {title} {\bibinfo {title} {{Measurement of the cross-section for W boson
  production in association with b-jets in pp collisions at $\sqrt{s}$ = 7 TeV
  with the ATLAS detector}},\ }\href {https://doi.org/10.1007/JHEP06(2013)084}
  {\bibfield  {journal} {\bibinfo  {journal} {JHEP}\ }\textbf {\bibinfo
  {volume} {06}},\ \bibinfo {pages} {084}},\ \Eprint
  {https://arxiv.org/abs/1302.2929} {arXiv:1302.2929 [hep-ex]} \BibitemShut
  {NoStop}%
\bibitem [{\citenamefont {Chatrchyan}\ \emph {et~al.}(2014)\citenamefont
  {Chatrchyan} \emph {et~al.}}]{CMS:2013xis}%
  \BibitemOpen
  \bibfield  {author} {\bibinfo {author} {\bibfnamefont {S.}~\bibnamefont
  {Chatrchyan}} \emph {et~al.} (\bibinfo {collaboration} {CMS}),\ }\bibfield
  {title} {\bibinfo {title} {{Measurement of the Production Cross Section for a
  W Boson and Two b Jets in pp Collisions at $\sqrt{s}$=7 TeV}},\ }\href
  {https://doi.org/10.1016/j.physletb.2014.06.041} {\bibfield  {journal}
  {\bibinfo  {journal} {Phys. Lett. B}\ }\textbf {\bibinfo {volume} {735}},\
  \bibinfo {pages} {204} (\bibinfo {year} {2014})},\ \Eprint
  {https://arxiv.org/abs/1312.6608} {arXiv:1312.6608 [hep-ex]} \BibitemShut
  {NoStop}%
\bibitem [{\citenamefont {Khachatryan}\ \emph {et~al.}(2017)\citenamefont
  {Khachatryan} \emph {et~al.}}]{CMS:2016eha}%
  \BibitemOpen
  \bibfield  {author} {\bibinfo {author} {\bibfnamefont {V.}~\bibnamefont
  {Khachatryan}} \emph {et~al.} (\bibinfo {collaboration} {CMS}),\ }\bibfield
  {title} {\bibinfo {title} {{Measurement of the production cross section of a
  W boson in association with two b jets in pp collisions at $\sqrt{s} =
  8{\,\mathrm{{TeV}}} $}},\ }\href
  {https://doi.org/10.1140/epjc/s10052-016-4573-z} {\bibfield  {journal}
  {\bibinfo  {journal} {Eur. Phys. J. C}\ }\textbf {\bibinfo {volume} {77}},\
  \bibinfo {pages} {92} (\bibinfo {year} {2017})},\ \Eprint
  {https://arxiv.org/abs/1608.07561} {arXiv:1608.07561 [hep-ex]} \BibitemShut
  {NoStop}%
\bibitem [{\citenamefont {Ellis}\ and\ \citenamefont
  {Veseli}(1999)}]{Ellis:1998fv}%
  \BibitemOpen
  \bibfield  {author} {\bibinfo {author} {\bibfnamefont {R.~K.}\ \bibnamefont
  {Ellis}}\ and\ \bibinfo {author} {\bibfnamefont {S.}~\bibnamefont {Veseli}},\
  }\bibfield  {title} {\bibinfo {title} {{Strong radiative corrections to W b
  anti-b production in p anti-p collisions}},\ }\href
  {https://doi.org/10.1103/PhysRevD.60.011501} {\bibfield  {journal} {\bibinfo
  {journal} {Phys. Rev. D}\ }\textbf {\bibinfo {volume} {60}},\ \bibinfo
  {pages} {011501} (\bibinfo {year} {1999})},\ \Eprint
  {https://arxiv.org/abs/hep-ph/9810489} {arXiv:hep-ph/9810489} \BibitemShut
  {NoStop}%
\bibitem [{\citenamefont {Febres~Cordero}\ \emph {et~al.}(2006)\citenamefont
  {Febres~Cordero}, \citenamefont {Reina},\ and\ \citenamefont
  {Wackeroth}}]{FebresCordero:2006nvf}%
  \BibitemOpen
  \bibfield  {author} {\bibinfo {author} {\bibfnamefont {F.}~\bibnamefont
  {Febres~Cordero}}, \bibinfo {author} {\bibfnamefont {L.}~\bibnamefont
  {Reina}},\ and\ \bibinfo {author} {\bibfnamefont {D.}~\bibnamefont
  {Wackeroth}},\ }\bibfield  {title} {\bibinfo {title} {{NLO QCD corrections to
  W boson production with a massive b-quark jet pair at the Tevatron p anti-p
  collider}},\ }\href {https://doi.org/10.1103/PhysRevD.74.034007} {\bibfield
  {journal} {\bibinfo  {journal} {Phys. Rev. D}\ }\textbf {\bibinfo {volume}
  {74}},\ \bibinfo {pages} {034007} (\bibinfo {year} {2006})},\ \Eprint
  {https://arxiv.org/abs/hep-ph/0606102} {arXiv:hep-ph/0606102} \BibitemShut
  {NoStop}%
\bibitem [{\citenamefont {Febres~Cordero}\ \emph {et~al.}(2009)\citenamefont
  {Febres~Cordero}, \citenamefont {Reina},\ and\ \citenamefont
  {Wackeroth}}]{FebresCordero:2009xzo}%
  \BibitemOpen
  \bibfield  {author} {\bibinfo {author} {\bibfnamefont {F.}~\bibnamefont
  {Febres~Cordero}}, \bibinfo {author} {\bibfnamefont {L.}~\bibnamefont
  {Reina}},\ and\ \bibinfo {author} {\bibfnamefont {D.}~\bibnamefont
  {Wackeroth}},\ }\bibfield  {title} {\bibinfo {title} {{W- and Z-boson
  production with a massive bottom-quark pair at the Large Hadron Collider}},\
  }\href {https://doi.org/10.1103/PhysRevD.80.034015} {\bibfield  {journal}
  {\bibinfo  {journal} {Phys. Rev. D}\ }\textbf {\bibinfo {volume} {80}},\
  \bibinfo {pages} {034015} (\bibinfo {year} {2009})},\ \Eprint
  {https://arxiv.org/abs/0906.1923} {arXiv:0906.1923 [hep-ph]} \BibitemShut
  {NoStop}%
\bibitem [{\citenamefont {Badger}\ \emph {et~al.}(2011)\citenamefont {Badger},
  \citenamefont {Campbell},\ and\ \citenamefont {Ellis}}]{Badger:2010mg}%
  \BibitemOpen
  \bibfield  {author} {\bibinfo {author} {\bibfnamefont {S.}~\bibnamefont
  {Badger}}, \bibinfo {author} {\bibfnamefont {J.~M.}\ \bibnamefont
  {Campbell}},\ and\ \bibinfo {author} {\bibfnamefont {R.~K.}\ \bibnamefont
  {Ellis}},\ }\bibfield  {title} {\bibinfo {title} {{QCD Corrections to the
  Hadronic Production of a Heavy Quark Pair and a W-Boson Including Decay
  Correlations}},\ }\href {https://doi.org/10.1007/JHEP03(2011)027} {\bibfield
  {journal} {\bibinfo  {journal} {JHEP}\ }\textbf {\bibinfo {volume} {03}},\
  \bibinfo {pages} {027}},\ \Eprint {https://arxiv.org/abs/1011.6647}
  {arXiv:1011.6647 [hep-ph]} \BibitemShut {NoStop}%
\bibitem [{\citenamefont {Oleari}\ and\ \citenamefont
  {Reina}(2011)}]{Oleari:2011ey}%
  \BibitemOpen
  \bibfield  {author} {\bibinfo {author} {\bibfnamefont {C.}~\bibnamefont
  {Oleari}}\ and\ \bibinfo {author} {\bibfnamefont {L.}~\bibnamefont {Reina}},\
  }\bibfield  {title} {\bibinfo {title} {{W +- b $\bar{b}$ production in
  POWHEG}},\ }\href {https://doi.org/10.1007/JHEP11(2011)040} {\bibfield
  {journal} {\bibinfo  {journal} {JHEP}\ }\textbf {\bibinfo {volume} {08}},\
  \bibinfo {pages} {061}},\ \bibinfo {note} {[Erratum: JHEP 11, 040 (2011)]},\
  \Eprint {https://arxiv.org/abs/1105.4488} {arXiv:1105.4488 [hep-ph]}
  \BibitemShut {NoStop}%
\bibitem [{\citenamefont {Frederix}\ \emph {et~al.}(2011)\citenamefont
  {Frederix}, \citenamefont {Frixione}, \citenamefont {Hirschi}, \citenamefont
  {Maltoni}, \citenamefont {Pittau},\ and\ \citenamefont
  {Torrielli}}]{Frederix:2011qg}%
  \BibitemOpen
  \bibfield  {author} {\bibinfo {author} {\bibfnamefont {R.}~\bibnamefont
  {Frederix}}, \bibinfo {author} {\bibfnamefont {S.}~\bibnamefont {Frixione}},
  \bibinfo {author} {\bibfnamefont {V.}~\bibnamefont {Hirschi}}, \bibinfo
  {author} {\bibfnamefont {F.}~\bibnamefont {Maltoni}}, \bibinfo {author}
  {\bibfnamefont {R.}~\bibnamefont {Pittau}},\ and\ \bibinfo {author}
  {\bibfnamefont {P.}~\bibnamefont {Torrielli}},\ }\bibfield  {title} {\bibinfo
  {title} {{W and $Z/\gamma*$ boson production in association with a
  bottom-antibottom pair}},\ }\href {https://doi.org/10.1007/JHEP09(2011)061}
  {\bibfield  {journal} {\bibinfo  {journal} {JHEP}\ }\textbf {\bibinfo
  {volume} {09}},\ \bibinfo {pages} {061}},\ \Eprint
  {https://arxiv.org/abs/1106.6019} {arXiv:1106.6019 [hep-ph]} \BibitemShut
  {NoStop}%
\bibitem [{\citenamefont {Luisoni}\ \emph {et~al.}(2015)\citenamefont
  {Luisoni}, \citenamefont {Oleari},\ and\ \citenamefont
  {Tramontano}}]{Luisoni:2015mpa}%
  \BibitemOpen
  \bibfield  {author} {\bibinfo {author} {\bibfnamefont {G.}~\bibnamefont
  {Luisoni}}, \bibinfo {author} {\bibfnamefont {C.}~\bibnamefont {Oleari}},\
  and\ \bibinfo {author} {\bibfnamefont {F.}~\bibnamefont {Tramontano}},\
  }\bibfield  {title} {\bibinfo {title} {{$ Wb\overline{b}j $ production at NLO
  with POWHEG+MiNLO}},\ }\href {https://doi.org/10.1007/JHEP04(2015)161}
  {\bibfield  {journal} {\bibinfo  {journal} {JHEP}\ }\textbf {\bibinfo
  {volume} {04}},\ \bibinfo {pages} {161}},\ \Eprint
  {https://arxiv.org/abs/1502.01213} {arXiv:1502.01213 [hep-ph]} \BibitemShut
  {NoStop}%
\bibitem [{\citenamefont {Anger}\ \emph {et~al.}(2018)\citenamefont {Anger},
  \citenamefont {Febres~Cordero}, \citenamefont {Ita},\ and\ \citenamefont
  {Sotnikov}}]{Anger:2017glm}%
  \BibitemOpen
  \bibfield  {author} {\bibinfo {author} {\bibfnamefont {F.~R.}\ \bibnamefont
  {Anger}}, \bibinfo {author} {\bibfnamefont {F.}~\bibnamefont
  {Febres~Cordero}}, \bibinfo {author} {\bibfnamefont {H.}~\bibnamefont
  {Ita}},\ and\ \bibinfo {author} {\bibfnamefont {V.}~\bibnamefont
  {Sotnikov}},\ }\bibfield  {title} {\bibinfo {title} {{NLO QCD predictions for
  $Wb\bar b$ production in association with up to three light jets at the
  LHC}},\ }\href {https://doi.org/10.1103/PhysRevD.97.036018} {\bibfield
  {journal} {\bibinfo  {journal} {Phys. Rev. D}\ }\textbf {\bibinfo {volume}
  {97}},\ \bibinfo {pages} {036018} (\bibinfo {year} {2018})},\ \Eprint
  {https://arxiv.org/abs/1712.05721} {arXiv:1712.05721 [hep-ph]} \BibitemShut
  {NoStop}%
\bibitem [{\citenamefont {Chawdhry}\ \emph {et~al.}(2020)\citenamefont
  {Chawdhry}, \citenamefont {Czakon}, \citenamefont {Mitov},\ and\
  \citenamefont {Poncelet}}]{Chawdhry:2019bji}%
  \BibitemOpen
  \bibfield  {author} {\bibinfo {author} {\bibfnamefont {H.~A.}\ \bibnamefont
  {Chawdhry}}, \bibinfo {author} {\bibfnamefont {M.~L.}\ \bibnamefont
  {Czakon}}, \bibinfo {author} {\bibfnamefont {A.}~\bibnamefont {Mitov}},\ and\
  \bibinfo {author} {\bibfnamefont {R.}~\bibnamefont {Poncelet}},\ }\bibfield
  {title} {\bibinfo {title} {{NNLO QCD corrections to three-photon production
  at the LHC}},\ }\href {https://doi.org/10.1007/JHEP02(2020)057} {\bibfield
  {journal} {\bibinfo  {journal} {JHEP}\ }\textbf {\bibinfo {volume} {02}},\
  \bibinfo {pages} {057}},\ \Eprint {https://arxiv.org/abs/1911.00479}
  {arXiv:1911.00479 [hep-ph]} \BibitemShut {NoStop}%
\bibitem [{\citenamefont {Kallweit}\ \emph {et~al.}(2021)\citenamefont
  {Kallweit}, \citenamefont {Sotnikov},\ and\ \citenamefont
  {Wiesemann}}]{Kallweit:2020gcp}%
  \BibitemOpen
  \bibfield  {author} {\bibinfo {author} {\bibfnamefont {S.}~\bibnamefont
  {Kallweit}}, \bibinfo {author} {\bibfnamefont {V.}~\bibnamefont {Sotnikov}},\
  and\ \bibinfo {author} {\bibfnamefont {M.}~\bibnamefont {Wiesemann}},\
  }\bibfield  {title} {\bibinfo {title} {{Triphoton production at hadron
  colliders in NNLO QCD}},\ }\href
  {https://doi.org/10.1016/j.physletb.2020.136013} {\bibfield  {journal}
  {\bibinfo  {journal} {Phys. Lett. B}\ }\textbf {\bibinfo {volume} {812}},\
  \bibinfo {pages} {136013} (\bibinfo {year} {2021})},\ \Eprint
  {https://arxiv.org/abs/2010.04681} {arXiv:2010.04681 [hep-ph]} \BibitemShut
  {NoStop}%
\bibitem [{\citenamefont {Chawdhry}\ \emph {et~al.}(2021)\citenamefont
  {Chawdhry}, \citenamefont {Czakon}, \citenamefont {Mitov},\ and\
  \citenamefont {Poncelet}}]{Chawdhry:2021hkp}%
  \BibitemOpen
  \bibfield  {author} {\bibinfo {author} {\bibfnamefont {H.~A.}\ \bibnamefont
  {Chawdhry}}, \bibinfo {author} {\bibfnamefont {M.}~\bibnamefont {Czakon}},
  \bibinfo {author} {\bibfnamefont {A.}~\bibnamefont {Mitov}},\ and\ \bibinfo
  {author} {\bibfnamefont {R.}~\bibnamefont {Poncelet}},\ }\bibfield  {title}
  {\bibinfo {title} {{NNLO QCD corrections to diphoton production with an
  additional jet at the LHC}},\ }\href
  {https://doi.org/10.1007/JHEP09(2021)093} {\bibfield  {journal} {\bibinfo
  {journal} {JHEP}\ }\textbf {\bibinfo {volume} {09}},\ \bibinfo {pages}
  {093}},\ \Eprint {https://arxiv.org/abs/2105.06940} {arXiv:2105.06940
  [hep-ph]} \BibitemShut {NoStop}%
\bibitem [{\citenamefont {Czakon}\ \emph {et~al.}(2021)\citenamefont {Czakon},
  \citenamefont {Mitov},\ and\ \citenamefont {Poncelet}}]{Czakon:2021mjy}%
  \BibitemOpen
  \bibfield  {author} {\bibinfo {author} {\bibfnamefont {M.}~\bibnamefont
  {Czakon}}, \bibinfo {author} {\bibfnamefont {A.}~\bibnamefont {Mitov}},\ and\
  \bibinfo {author} {\bibfnamefont {R.}~\bibnamefont {Poncelet}},\ }\bibfield
  {title} {\bibinfo {title} {{Next-to-Next-to-Leading Order Study of Three-Jet
  Production at the LHC}},\ }\href
  {https://doi.org/10.1103/PhysRevLett.127.152001} {\bibfield  {journal}
  {\bibinfo  {journal} {Phys. Rev. Lett.}\ }\textbf {\bibinfo {volume} {127}},\
  \bibinfo {pages} {152001} (\bibinfo {year} {2021})},\ \Eprint
  {https://arxiv.org/abs/2106.05331} {arXiv:2106.05331 [hep-ph]} \BibitemShut
  {NoStop}%
\bibitem [{\citenamefont {Badger}\ \emph
  {et~al.}(2022{\natexlab{a}})\citenamefont {Badger}, \citenamefont {Gehrmann},
  \citenamefont {Marcoli},\ and\ \citenamefont {Moodie}}]{Badger:2021ohm}%
  \BibitemOpen
  \bibfield  {author} {\bibinfo {author} {\bibfnamefont {S.}~\bibnamefont
  {Badger}}, \bibinfo {author} {\bibfnamefont {T.}~\bibnamefont {Gehrmann}},
  \bibinfo {author} {\bibfnamefont {M.}~\bibnamefont {Marcoli}},\ and\ \bibinfo
  {author} {\bibfnamefont {R.}~\bibnamefont {Moodie}},\ }\bibfield  {title}
  {\bibinfo {title} {{Next-to-leading order QCD corrections to
  diphoton-plus-jet production through gluon fusion at the LHC}},\ }\href
  {https://doi.org/10.1016/j.physletb.2021.136802} {\bibfield  {journal}
  {\bibinfo  {journal} {Phys. Lett. B}\ }\textbf {\bibinfo {volume} {824}},\
  \bibinfo {pages} {136802} (\bibinfo {year} {2022}{\natexlab{a}})},\ \Eprint
  {https://arxiv.org/abs/2109.12003} {arXiv:2109.12003 [hep-ph]} \BibitemShut
  {NoStop}%
\bibitem [{\citenamefont {Chen}\ \emph {et~al.}(2022)\citenamefont {Chen},
  \citenamefont {Gehrmann}, \citenamefont {Glover}, \citenamefont {Huss},\ and\
  \citenamefont {Marcoli}}]{Chen:2022ktf}%
  \BibitemOpen
  \bibfield  {author} {\bibinfo {author} {\bibfnamefont {X.}~\bibnamefont
  {Chen}}, \bibinfo {author} {\bibfnamefont {T.}~\bibnamefont {Gehrmann}},
  \bibinfo {author} {\bibfnamefont {N.}~\bibnamefont {Glover}}, \bibinfo
  {author} {\bibfnamefont {A.}~\bibnamefont {Huss}},\ and\ \bibinfo {author}
  {\bibfnamefont {M.}~\bibnamefont {Marcoli}},\ }\bibfield  {title} {\bibinfo
  {title} {{Automation of antenna subtraction in colour space: gluonic
  processes}},\ }\href@noop {} {\  (\bibinfo {year} {2022})},\ \Eprint
  {https://arxiv.org/abs/2203.13531} {arXiv:2203.13531 [hep-ph]} \BibitemShut
  {NoStop}%
\bibitem [{\citenamefont {Papadopoulos}\ \emph {et~al.}(2016)\citenamefont
  {Papadopoulos}, \citenamefont {Tommasini},\ and\ \citenamefont
  {Wever}}]{Papadopoulos:2015jft}%
  \BibitemOpen
  \bibfield  {author} {\bibinfo {author} {\bibfnamefont {C.~G.}\ \bibnamefont
  {Papadopoulos}}, \bibinfo {author} {\bibfnamefont {D.}~\bibnamefont
  {Tommasini}},\ and\ \bibinfo {author} {\bibfnamefont {C.}~\bibnamefont
  {Wever}},\ }\bibfield  {title} {\bibinfo {title} {{The Pentabox Master
  Integrals with the Simplified Differential Equations approach}},\ }\href
  {https://doi.org/10.1007/JHEP04(2016)078} {\bibfield  {journal} {\bibinfo
  {journal} {JHEP}\ }\textbf {\bibinfo {volume} {04}},\ \bibinfo {pages}
  {078}},\ \Eprint {https://arxiv.org/abs/1511.09404} {arXiv:1511.09404
  [hep-ph]} \BibitemShut {NoStop}%
\bibitem [{\citenamefont {Abreu}\ \emph {et~al.}(2020)\citenamefont {Abreu},
  \citenamefont {Ita}, \citenamefont {Moriello}, \citenamefont {Page},
  \citenamefont {Tschernow},\ and\ \citenamefont {Zeng}}]{Abreu:2020jxa}%
  \BibitemOpen
  \bibfield  {author} {\bibinfo {author} {\bibfnamefont {S.}~\bibnamefont
  {Abreu}}, \bibinfo {author} {\bibfnamefont {H.}~\bibnamefont {Ita}}, \bibinfo
  {author} {\bibfnamefont {F.}~\bibnamefont {Moriello}}, \bibinfo {author}
  {\bibfnamefont {B.}~\bibnamefont {Page}}, \bibinfo {author} {\bibfnamefont
  {W.}~\bibnamefont {Tschernow}},\ and\ \bibinfo {author} {\bibfnamefont
  {M.}~\bibnamefont {Zeng}},\ }\bibfield  {title} {\bibinfo {title} {{Two-Loop
  Integrals for Planar Five-Point One-Mass Processes}},\ }\href
  {https://doi.org/10.1007/JHEP11(2020)117} {\bibfield  {journal} {\bibinfo
  {journal} {JHEP}\ }\textbf {\bibinfo {volume} {11}},\ \bibinfo {pages}
  {117}},\ \Eprint {https://arxiv.org/abs/2005.04195} {arXiv:2005.04195
  [hep-ph]} \BibitemShut {NoStop}%
\bibitem [{\citenamefont {Canko}\ \emph {et~al.}(2021)\citenamefont {Canko},
  \citenamefont {Papadopoulos},\ and\ \citenamefont
  {Syrrakos}}]{Canko:2020ylt}%
  \BibitemOpen
  \bibfield  {author} {\bibinfo {author} {\bibfnamefont {D.~D.}\ \bibnamefont
  {Canko}}, \bibinfo {author} {\bibfnamefont {C.~G.}\ \bibnamefont
  {Papadopoulos}},\ and\ \bibinfo {author} {\bibfnamefont {N.}~\bibnamefont
  {Syrrakos}},\ }\bibfield  {title} {\bibinfo {title} {{Analytic representation
  of all planar two-loop five-point Master Integrals with one off-shell leg}},\
  }\href {https://doi.org/10.1007/JHEP01(2021)199} {\bibfield  {journal}
  {\bibinfo  {journal} {JHEP}\ }\textbf {\bibinfo {volume} {01}},\ \bibinfo
  {pages} {199}},\ \Eprint {https://arxiv.org/abs/2009.13917} {arXiv:2009.13917
  [hep-ph]} \BibitemShut {NoStop}%
\bibitem [{\citenamefont {Syrrakos}(2021)}]{Syrrakos:2020kba}%
  \BibitemOpen
  \bibfield  {author} {\bibinfo {author} {\bibfnamefont {N.}~\bibnamefont
  {Syrrakos}},\ }\bibfield  {title} {\bibinfo {title} {{Pentagon integrals to
  arbitrary order in the dimensional regulator}},\ }\href
  {https://doi.org/10.1007/JHEP06(2021)037} {\bibfield  {journal} {\bibinfo
  {journal} {JHEP}\ }\textbf {\bibinfo {volume} {06}},\ \bibinfo {pages}
  {037}},\ \Eprint {https://arxiv.org/abs/2012.10635} {arXiv:2012.10635
  [hep-ph]} \BibitemShut {NoStop}%
\bibitem [{\citenamefont {Badger}\ \emph
  {et~al.}(2021{\natexlab{a}})\citenamefont {Badger}, \citenamefont
  {Hartanto},\ and\ \citenamefont {Zoia}}]{Badger:2021nhg}%
  \BibitemOpen
  \bibfield  {author} {\bibinfo {author} {\bibfnamefont {S.}~\bibnamefont
  {Badger}}, \bibinfo {author} {\bibfnamefont {H.~B.}\ \bibnamefont
  {Hartanto}},\ and\ \bibinfo {author} {\bibfnamefont {S.}~\bibnamefont
  {Zoia}},\ }\bibfield  {title} {\bibinfo {title} {{Two-Loop QCD Corrections to
  $Wb\bar{b}$ Production at Hadron Colliders}},\ }\href
  {https://doi.org/10.1103/PhysRevLett.127.012001} {\bibfield  {journal}
  {\bibinfo  {journal} {Phys. Rev. Lett.}\ }\textbf {\bibinfo {volume} {127}},\
  \bibinfo {pages} {012001} (\bibinfo {year} {2021}{\natexlab{a}})},\ \Eprint
  {https://arxiv.org/abs/2102.02516} {arXiv:2102.02516 [hep-ph]} \BibitemShut
  {NoStop}%
\bibitem [{\citenamefont {Chicherin}\ \emph {et~al.}(2021)\citenamefont
  {Chicherin}, \citenamefont {Sotnikov},\ and\ \citenamefont
  {Zoia}}]{Chicherin:2021dyp}%
  \BibitemOpen
  \bibfield  {author} {\bibinfo {author} {\bibfnamefont {D.}~\bibnamefont
  {Chicherin}}, \bibinfo {author} {\bibfnamefont {V.}~\bibnamefont
  {Sotnikov}},\ and\ \bibinfo {author} {\bibfnamefont {S.}~\bibnamefont
  {Zoia}},\ }\bibfield  {title} {\bibinfo {title} {{Pentagon Functions for
  One-Mass Planar Scattering Amplitudes}},\ }\href@noop {} {\  (\bibinfo {year}
  {2021})},\ \Eprint {https://arxiv.org/abs/2110.10111} {arXiv:2110.10111
  [hep-ph]} \BibitemShut {NoStop}%
\bibitem [{\citenamefont {Papadopoulos}\ and\ \citenamefont
  {Wever}(2020)}]{Papadopoulos:2019iam}%
  \BibitemOpen
  \bibfield  {author} {\bibinfo {author} {\bibfnamefont {C.~G.}\ \bibnamefont
  {Papadopoulos}}\ and\ \bibinfo {author} {\bibfnamefont {C.}~\bibnamefont
  {Wever}},\ }\bibfield  {title} {\bibinfo {title} {{Internal Reduction method
  for computing Feynman Integrals}},\ }\href
  {https://doi.org/10.1007/JHEP02(2020)112} {\bibfield  {journal} {\bibinfo
  {journal} {JHEP}\ }\textbf {\bibinfo {volume} {02}},\ \bibinfo {pages}
  {112}},\ \Eprint {https://arxiv.org/abs/1910.06275} {arXiv:1910.06275
  [hep-ph]} \BibitemShut {NoStop}%
\bibitem [{\citenamefont {Abreu}\ \emph
  {et~al.}(2022{\natexlab{a}})\citenamefont {Abreu}, \citenamefont {Ita},
  \citenamefont {Page},\ and\ \citenamefont {Tschernow}}]{Abreu:2021smk}%
  \BibitemOpen
  \bibfield  {author} {\bibinfo {author} {\bibfnamefont {S.}~\bibnamefont
  {Abreu}}, \bibinfo {author} {\bibfnamefont {H.}~\bibnamefont {Ita}}, \bibinfo
  {author} {\bibfnamefont {B.}~\bibnamefont {Page}},\ and\ \bibinfo {author}
  {\bibfnamefont {W.}~\bibnamefont {Tschernow}},\ }\bibfield  {title} {\bibinfo
  {title} {{Two-loop hexa-box integrals for non-planar five-point one-mass
  processes}},\ }\href {https://doi.org/10.1007/JHEP03(2022)182} {\bibfield
  {journal} {\bibinfo  {journal} {JHEP}\ }\textbf {\bibinfo {volume} {03}},\
  \bibinfo {pages} {182}},\ \Eprint {https://arxiv.org/abs/2107.14180}
  {arXiv:2107.14180 [hep-ph]} \BibitemShut {NoStop}%
\bibitem [{\citenamefont {Kardos}\ \emph {et~al.}(2022)\citenamefont {Kardos},
  \citenamefont {Papadopoulos}, \citenamefont {Smirnov}, \citenamefont
  {Syrrakos},\ and\ \citenamefont {Wever}}]{Kardos:2022tpo}%
  \BibitemOpen
  \bibfield  {author} {\bibinfo {author} {\bibfnamefont {A.}~\bibnamefont
  {Kardos}}, \bibinfo {author} {\bibfnamefont {C.~G.}\ \bibnamefont
  {Papadopoulos}}, \bibinfo {author} {\bibfnamefont {A.~V.}\ \bibnamefont
  {Smirnov}}, \bibinfo {author} {\bibfnamefont {N.}~\bibnamefont {Syrrakos}},\
  and\ \bibinfo {author} {\bibfnamefont {C.}~\bibnamefont {Wever}},\ }\bibfield
   {title} {\bibinfo {title} {{Two-loop non-planar hexa-box integrals with one
  massive leg}},\ }\href@noop {} {\  (\bibinfo {year} {2022})},\ \Eprint
  {https://arxiv.org/abs/2201.07509} {arXiv:2201.07509 [hep-ph]} \BibitemShut
  {NoStop}%
\bibitem [{\citenamefont {Badger}\ \emph
  {et~al.}(2021{\natexlab{b}})\citenamefont {Badger}, \citenamefont {Hartanto},
  \citenamefont {Kry\'s},\ and\ \citenamefont {Zoia}}]{Badger:2021ega}%
  \BibitemOpen
  \bibfield  {author} {\bibinfo {author} {\bibfnamefont {S.}~\bibnamefont
  {Badger}}, \bibinfo {author} {\bibfnamefont {H.~B.}\ \bibnamefont
  {Hartanto}}, \bibinfo {author} {\bibfnamefont {J.}~\bibnamefont {Kry\'s}},\
  and\ \bibinfo {author} {\bibfnamefont {S.}~\bibnamefont {Zoia}},\ }\bibfield
  {title} {\bibinfo {title} {{Two-loop leading-colour QCD helicity amplitudes
  for Higgs boson production in association with a bottom-quark pair at the
  LHC}},\ }\href {https://doi.org/10.1007/JHEP11(2021)012} {\bibfield
  {journal} {\bibinfo  {journal} {JHEP}\ }\textbf {\bibinfo {volume} {11}},\
  \bibinfo {pages} {012}},\ \Eprint {https://arxiv.org/abs/2107.14733}
  {arXiv:2107.14733 [hep-ph]} \BibitemShut {NoStop}%
\bibitem [{\citenamefont {Abreu}\ \emph
  {et~al.}(2022{\natexlab{b}})\citenamefont {Abreu}, \citenamefont
  {Febres~Cordero}, \citenamefont {Ita}, \citenamefont {Klinkert},
  \citenamefont {Page},\ and\ \citenamefont {Sotnikov}}]{Abreu:2021asb}%
  \BibitemOpen
  \bibfield  {author} {\bibinfo {author} {\bibfnamefont {S.}~\bibnamefont
  {Abreu}}, \bibinfo {author} {\bibfnamefont {F.}~\bibnamefont
  {Febres~Cordero}}, \bibinfo {author} {\bibfnamefont {H.}~\bibnamefont {Ita}},
  \bibinfo {author} {\bibfnamefont {M.}~\bibnamefont {Klinkert}}, \bibinfo
  {author} {\bibfnamefont {B.}~\bibnamefont {Page}},\ and\ \bibinfo {author}
  {\bibfnamefont {V.}~\bibnamefont {Sotnikov}},\ }\bibfield  {title} {\bibinfo
  {title} {{Leading-color two-loop amplitudes for four partons and a W boson in
  QCD}},\ }\href {https://doi.org/10.1007/JHEP04(2022)042} {\bibfield
  {journal} {\bibinfo  {journal} {JHEP}\ }\textbf {\bibinfo {volume} {04}},\
  \bibinfo {pages} {042}},\ \Eprint {https://arxiv.org/abs/2110.07541}
  {arXiv:2110.07541 [hep-ph]} \BibitemShut {NoStop}%
\bibitem [{\citenamefont {Badger}\ \emph
  {et~al.}(2022{\natexlab{b}})\citenamefont {Badger}, \citenamefont {Hartanto},
  \citenamefont {Kry\'s},\ and\ \citenamefont {Zoia}}]{Badger:2022ncb}%
  \BibitemOpen
  \bibfield  {author} {\bibinfo {author} {\bibfnamefont {S.}~\bibnamefont
  {Badger}}, \bibinfo {author} {\bibfnamefont {H.~B.}\ \bibnamefont
  {Hartanto}}, \bibinfo {author} {\bibfnamefont {J.}~\bibnamefont {Kry\'s}},\
  and\ \bibinfo {author} {\bibfnamefont {S.}~\bibnamefont {Zoia}},\ }\bibfield
  {title} {\bibinfo {title} {{Two-loop leading colour helicity amplitudes for
  $W^\pm\gamma+j$ production at the LHC}},\ }\href@noop {} {\  (\bibinfo {year}
  {2022}{\natexlab{b}})},\ \Eprint {https://arxiv.org/abs/2201.04075}
  {arXiv:2201.04075 [hep-ph]} \BibitemShut {NoStop}%
\bibitem [{\citenamefont {Czakon}(2010)}]{Czakon:2010td}%
  \BibitemOpen
  \bibfield  {author} {\bibinfo {author} {\bibfnamefont {M.}~\bibnamefont
  {Czakon}},\ }\bibfield  {title} {\bibinfo {title} {{A novel subtraction
  scheme for double-real radiation at NNLO}},\ }\href
  {https://doi.org/10.1016/j.physletb.2010.08.036} {\bibfield  {journal}
  {\bibinfo  {journal} {Phys. Lett. B}\ }\textbf {\bibinfo {volume} {693}},\
  \bibinfo {pages} {259} (\bibinfo {year} {2010})},\ \Eprint
  {https://arxiv.org/abs/1005.0274} {arXiv:1005.0274 [hep-ph]} \BibitemShut
  {NoStop}%
\bibitem [{\citenamefont {Czakon}\ and\ \citenamefont
  {Heymes}(2014)}]{Czakon:2014oma}%
  \BibitemOpen
  \bibfield  {author} {\bibinfo {author} {\bibfnamefont {M.}~\bibnamefont
  {Czakon}}\ and\ \bibinfo {author} {\bibfnamefont {D.}~\bibnamefont
  {Heymes}},\ }\bibfield  {title} {\bibinfo {title} {{Four-dimensional
  formulation of the sector-improved residue subtraction scheme}},\ }\href
  {https://doi.org/10.1016/j.nuclphysb.2014.11.006} {\bibfield  {journal}
  {\bibinfo  {journal} {Nucl. Phys. B}\ }\textbf {\bibinfo {volume} {890}},\
  \bibinfo {pages} {152} (\bibinfo {year} {2014})},\ \Eprint
  {https://arxiv.org/abs/1408.2500} {arXiv:1408.2500 [hep-ph]} \BibitemShut
  {NoStop}%
\bibitem [{\citenamefont {Czakon}\ \emph {et~al.}(2019)\citenamefont {Czakon},
  \citenamefont {van Hameren}, \citenamefont {Mitov},\ and\ \citenamefont
  {Poncelet}}]{Czakon:2019tmo}%
  \BibitemOpen
  \bibfield  {author} {\bibinfo {author} {\bibfnamefont {M.}~\bibnamefont
  {Czakon}}, \bibinfo {author} {\bibfnamefont {A.}~\bibnamefont {van Hameren}},
  \bibinfo {author} {\bibfnamefont {A.}~\bibnamefont {Mitov}},\ and\ \bibinfo
  {author} {\bibfnamefont {R.}~\bibnamefont {Poncelet}},\ }\bibfield  {title}
  {\bibinfo {title} {{Single-jet inclusive rates with exact color at $
  \mathcal{O} $ ($ {\alpha}_s^4 $)}},\ }\href
  {https://doi.org/10.1007/JHEP10(2019)262} {\bibfield  {journal} {\bibinfo
  {journal} {JHEP}\ }\textbf {\bibinfo {volume} {10}},\ \bibinfo {pages}
  {262}},\ \Eprint {https://arxiv.org/abs/1907.12911} {arXiv:1907.12911
  [hep-ph]} \BibitemShut {NoStop}%
\bibitem [{\citenamefont {Bury}\ and\ \citenamefont {van
  Hameren}(2015)}]{Bury:2015dla}%
  \BibitemOpen
  \bibfield  {author} {\bibinfo {author} {\bibfnamefont {M.}~\bibnamefont
  {Bury}}\ and\ \bibinfo {author} {\bibfnamefont {A.}~\bibnamefont {van
  Hameren}},\ }\bibfield  {title} {\bibinfo {title} {{Numerical evaluation of
  multi-gluon amplitudes for High Energy Factorization}},\ }\href
  {https://doi.org/10.1016/j.cpc.2015.06.023} {\bibfield  {journal} {\bibinfo
  {journal} {Comput. Phys. Commun.}\ }\textbf {\bibinfo {volume} {196}},\
  \bibinfo {pages} {592} (\bibinfo {year} {2015})},\ \Eprint
  {https://arxiv.org/abs/1503.08612} {arXiv:1503.08612 [hep-ph]} \BibitemShut
  {NoStop}%
\bibitem [{\citenamefont {Buccioni}\ \emph {et~al.}(2018)\citenamefont
  {Buccioni}, \citenamefont {Pozzorini},\ and\ \citenamefont
  {Zoller}}]{Buccioni:2017yxi}%
  \BibitemOpen
  \bibfield  {author} {\bibinfo {author} {\bibfnamefont {F.}~\bibnamefont
  {Buccioni}}, \bibinfo {author} {\bibfnamefont {S.}~\bibnamefont
  {Pozzorini}},\ and\ \bibinfo {author} {\bibfnamefont {M.}~\bibnamefont
  {Zoller}},\ }\bibfield  {title} {\bibinfo {title} {{On-the-fly reduction of
  open loops}},\ }\href {https://doi.org/10.1140/epjc/s10052-018-5562-1}
  {\bibfield  {journal} {\bibinfo  {journal} {Eur. Phys. J. C}\ }\textbf
  {\bibinfo {volume} {78}},\ \bibinfo {pages} {70} (\bibinfo {year} {2018})},\
  \Eprint {https://arxiv.org/abs/1710.11452} {arXiv:1710.11452 [hep-ph]}
  \BibitemShut {NoStop}%
\bibitem [{\citenamefont {Buccioni}\ \emph {et~al.}(2019)\citenamefont
  {Buccioni}, \citenamefont {Lang}, \citenamefont {Lindert}, \citenamefont
  {Maierh\"ofer}, \citenamefont {Pozzorini}, \citenamefont {Zhang},\ and\
  \citenamefont {Zoller}}]{Buccioni:2019sur}%
  \BibitemOpen
  \bibfield  {author} {\bibinfo {author} {\bibfnamefont {F.}~\bibnamefont
  {Buccioni}}, \bibinfo {author} {\bibfnamefont {J.-N.}\ \bibnamefont {Lang}},
  \bibinfo {author} {\bibfnamefont {J.~M.}\ \bibnamefont {Lindert}}, \bibinfo
  {author} {\bibfnamefont {P.}~\bibnamefont {Maierh\"ofer}}, \bibinfo {author}
  {\bibfnamefont {S.}~\bibnamefont {Pozzorini}}, \bibinfo {author}
  {\bibfnamefont {H.}~\bibnamefont {Zhang}},\ and\ \bibinfo {author}
  {\bibfnamefont {M.~F.}\ \bibnamefont {Zoller}},\ }\bibfield  {title}
  {\bibinfo {title} {{OpenLoops 2}},\ }\href
  {https://doi.org/10.1140/epjc/s10052-019-7306-2} {\bibfield  {journal}
  {\bibinfo  {journal} {Eur. Phys. J. C}\ }\textbf {\bibinfo {volume} {79}},\
  \bibinfo {pages} {866} (\bibinfo {year} {2019})},\ \Eprint
  {https://arxiv.org/abs/1907.13071} {arXiv:1907.13071 [hep-ph]} \BibitemShut
  {NoStop}%
\bibitem [{\citenamefont {Chen}(2021)}]{Chen:2019wyb}%
  \BibitemOpen
  \bibfield  {author} {\bibinfo {author} {\bibfnamefont {L.}~\bibnamefont
  {Chen}},\ }\bibfield  {title} {\bibinfo {title} {{A prescription for
  projectors to compute helicity amplitudes in D dimensions}},\ }\href
  {https://doi.org/10.1140/epjc/s10052-021-09210-9} {\bibfield  {journal}
  {\bibinfo  {journal} {Eur. Phys. J. C}\ }\textbf {\bibinfo {volume} {81}},\
  \bibinfo {pages} {417} (\bibinfo {year} {2021})},\ \Eprint
  {https://arxiv.org/abs/1904.00705} {arXiv:1904.00705 [hep-ph]} \BibitemShut
  {NoStop}%
\bibitem [{\citenamefont {Peraro}\ and\ \citenamefont
  {Tancredi}(2021)}]{Peraro:2020sfm}%
  \BibitemOpen
  \bibfield  {author} {\bibinfo {author} {\bibfnamefont {T.}~\bibnamefont
  {Peraro}}\ and\ \bibinfo {author} {\bibfnamefont {L.}~\bibnamefont
  {Tancredi}},\ }\bibfield  {title} {\bibinfo {title} {{Tensor decomposition
  for bosonic and fermionic scattering amplitudes}},\ }\href
  {https://doi.org/10.1103/PhysRevD.103.054042} {\bibfield  {journal} {\bibinfo
   {journal} {Phys. Rev. D}\ }\textbf {\bibinfo {volume} {103}},\ \bibinfo
  {pages} {054042} (\bibinfo {year} {2021})},\ \Eprint
  {https://arxiv.org/abs/2012.00820} {arXiv:2012.00820 [hep-ph]} \BibitemShut
  {NoStop}%
\bibitem [{\citenamefont {Peraro}(2016)}]{Peraro:2016wsq}%
  \BibitemOpen
  \bibfield  {author} {\bibinfo {author} {\bibfnamefont {T.}~\bibnamefont
  {Peraro}},\ }\bibfield  {title} {\bibinfo {title} {{Scattering amplitudes
  over finite fields and multivariate functional reconstruction}},\ }\href
  {https://doi.org/10.1007/JHEP12(2016)030} {\bibfield  {journal} {\bibinfo
  {journal} {JHEP}\ }\textbf {\bibinfo {volume} {12}},\ \bibinfo {pages}
  {030}},\ \Eprint {https://arxiv.org/abs/1608.01902} {arXiv:1608.01902
  [hep-ph]} \BibitemShut {NoStop}%
\bibitem [{\citenamefont {Peraro}(2019)}]{Peraro:2019svx}%
  \BibitemOpen
  \bibfield  {author} {\bibinfo {author} {\bibfnamefont {T.}~\bibnamefont
  {Peraro}},\ }\bibfield  {title} {\bibinfo {title} {{FiniteFlow: multivariate
  functional reconstruction using finite fields and dataflow graphs}},\ }\href
  {https://doi.org/10.1007/JHEP07(2019)031} {\bibfield  {journal} {\bibinfo
  {journal} {JHEP}\ }\textbf {\bibinfo {volume} {07}},\ \bibinfo {pages}
  {031}},\ \Eprint {https://arxiv.org/abs/1905.08019} {arXiv:1905.08019
  [hep-ph]} \BibitemShut {NoStop}%
\bibitem [{\citenamefont {Heller}\ and\ \citenamefont {von
  Manteuffel}(2022)}]{Heller:2021qkz}%
  \BibitemOpen
  \bibfield  {author} {\bibinfo {author} {\bibfnamefont {M.}~\bibnamefont
  {Heller}}\ and\ \bibinfo {author} {\bibfnamefont {A.}~\bibnamefont {von
  Manteuffel}},\ }\bibfield  {title} {\bibinfo {title} {{MultivariateApart:
  Generalized partial fractions}},\ }\href
  {https://doi.org/10.1016/j.cpc.2021.108174} {\bibfield  {journal} {\bibinfo
  {journal} {Comput. Phys. Commun.}\ }\textbf {\bibinfo {volume} {271}},\
  \bibinfo {pages} {108174} (\bibinfo {year} {2022})},\ \Eprint
  {https://arxiv.org/abs/2101.08283} {arXiv:2101.08283 [cs.SC]} \BibitemShut
  {NoStop}%
\bibitem [{\citenamefont {Decker}\ \emph {et~al.}(2021)\citenamefont {Decker},
  \citenamefont {Greuel}, \citenamefont {Pfister},\ and\ \citenamefont
  {Sch\"onemann}}]{DGPS}%
  \BibitemOpen
  \bibfield  {author} {\bibinfo {author} {\bibfnamefont {W.}~\bibnamefont
  {Decker}}, \bibinfo {author} {\bibfnamefont {G.-M.}\ \bibnamefont {Greuel}},
  \bibinfo {author} {\bibfnamefont {G.}~\bibnamefont {Pfister}},\ and\ \bibinfo
  {author} {\bibfnamefont {H.}~\bibnamefont {Sch\"onemann}},\ }\href@noop {}
  {\bibinfo {title} {{\sc Singular} {4-2-1} --- {A} computer algebra system for
  polynomial computations}},\ \bibinfo {howpublished}
  {\href{http://www.singular.uni-kl.de}{http://www.singular.uni-kl.de}}
  (\bibinfo {year} {2021})\BibitemShut {NoStop}%
\bibitem [{\citenamefont {Ball}\ \emph {et~al.}(2017)\citenamefont {Ball} \emph
  {et~al.}}]{NNPDF:2017mvq}%
  \BibitemOpen
  \bibfield  {author} {\bibinfo {author} {\bibfnamefont {R.~D.}\ \bibnamefont
  {Ball}} \emph {et~al.} (\bibinfo {collaboration} {NNPDF}),\ }\bibfield
  {title} {\bibinfo {title} {{Parton distributions from high-precision collider
  data}},\ }\href {https://doi.org/10.1140/epjc/s10052-017-5199-5} {\bibfield
  {journal} {\bibinfo  {journal} {Eur. Phys. J. C}\ }\textbf {\bibinfo {volume}
  {77}},\ \bibinfo {pages} {663} (\bibinfo {year} {2017})},\ \Eprint
  {https://arxiv.org/abs/1706.00428} {arXiv:1706.00428 [hep-ph]} \BibitemShut
  {NoStop}%
\bibitem [{\citenamefont {Banfi}\ \emph {et~al.}(2006)\citenamefont {Banfi},
  \citenamefont {Salam},\ and\ \citenamefont {Zanderighi}}]{Banfi:2006hf}%
  \BibitemOpen
  \bibfield  {author} {\bibinfo {author} {\bibfnamefont {A.}~\bibnamefont
  {Banfi}}, \bibinfo {author} {\bibfnamefont {G.~P.}\ \bibnamefont {Salam}},\
  and\ \bibinfo {author} {\bibfnamefont {G.}~\bibnamefont {Zanderighi}},\
  }\bibfield  {title} {\bibinfo {title} {{Infrared safe definition of jet
  flavor}},\ }\href {https://doi.org/10.1140/epjc/s2006-02552-4} {\bibfield
  {journal} {\bibinfo  {journal} {Eur. Phys. J. C}\ }\textbf {\bibinfo {volume}
  {47}},\ \bibinfo {pages} {113} (\bibinfo {year} {2006})},\ \Eprint
  {https://arxiv.org/abs/hep-ph/0601139} {arXiv:hep-ph/0601139} \BibitemShut
  {NoStop}%
\bibitem [{\citenamefont {Stewart}\ and\ \citenamefont
  {Tackmann}(2012)}]{Stewart:2011cf}%
  \BibitemOpen
  \bibfield  {author} {\bibinfo {author} {\bibfnamefont {I.~W.}\ \bibnamefont
  {Stewart}}\ and\ \bibinfo {author} {\bibfnamefont {F.~J.}\ \bibnamefont
  {Tackmann}},\ }\bibfield  {title} {\bibinfo {title} {{Theory Uncertainties
  for Higgs and Other Searches Using Jet Bins}},\ }\href
  {https://doi.org/10.1103/PhysRevD.85.034011} {\bibfield  {journal} {\bibinfo
  {journal} {Phys. Rev. D}\ }\textbf {\bibinfo {volume} {85}},\ \bibinfo
  {pages} {034011} (\bibinfo {year} {2012})},\ \Eprint
  {https://arxiv.org/abs/1107.2117} {arXiv:1107.2117 [hep-ph]} \BibitemShut
  {NoStop}%
\end{thebibliography}%

\end{document}